\documentclass[prl,twocolumn,letterpaper,byrevtex,floatfix,showpacs,%
preprintnumbers,superscriptaddress,amsfonts,amssymb,amsmath]{revtex4}

\usepackage{rcs}		

\usepackage{mathptmx}           
\usepackage[dvips]{graphicx}	
\usepackage{amsbsy}		
\usepackage{url}		
\usepackage{graphicx}		

\renewcommand*{\i}{\mathrm{i}}
\renewcommand*{\deg}{^\circ}

\newcommand*{\eg}{\emph{e.g.}}

\newcommand*{\ie}{\emph{i.e.}}

\newcommand*{\sub}[1]{_{\mathrm{#1}}}
\renewcommand*{\sup}[1]{^{\mathrm{#1}}}
\renewcommand*{\vec}[1]{\mathbf{#1}}

\newcommand*{\cross}{\boldsymbol{\times}}
\renewcommand{\d}[1]{\mathrm{d}#1}
\newcommand*{\eps}{\varepsilon}
\newcommand*{\epsz}{\eps_{0}}

\newcommand*{\E}{\vec{E}}
\newcommand*{\B}{\vec{B}}

\renewcommand*{\j}{\vec{j}}
\newcommand*{\J}{\vec{J}}
\newcommand*{\Jem}{\J\sup{em}}
\newcommand*{\Jzem}{J_z\sup{em}}
\newcommand*{\Jmech}{\J\sup{mech}}
\newcommand*{\Jtot}{\J\sup{tot}}
\renewcommand*{\k}{\vec{k}}
\newcommand*{\Lem}{\vec{L}\sup{em}}
\newcommand*{\Lzem}{L_z\sup{em}}

\newcommand*{\p}{\vec{p}}
\newcommand*{\pmech}{\p\sup{mech}}
\newcommand*{\Sem}{\vec{S}\sup{em}}
\newcommand*{\w}{\omega}
\newcommand*{\x}{\vec{x}}

\begin{document}

\title{Utilization of photon orbital angular momentum in the
low-frequency radio domain}

\author{B.\,Thid\'e}
 \altaffiliation[Also at ]{LOIS Space Centre, V\"axj\"o University,
 SE-351\,95 V\"axj\"o, Sweden}
 \email{bt@irfu.se}
 \affiliation{%
 Swedish Institute of Space Physics,
 \AA ngstr\"om Laboratory,
 P.\,O.\,Box 537,
 SE-751\,21 Uppsala,
 Sweden}%

\author{H.\,Then}
 \affiliation{%
 Institute of Physics, 
 Carl-von-Ossietzky Universit\"at Oldenburg,
 D-261\,11 Oldenburg,
 Germany}%

\author{J.\,Sj\"oholm}
\author{K.\,Palmer}
 \affiliation{%
 Department of Astronomy and Space Physics,
 \AA ngstr\"om Laboratory,
 P.\,O.\,Box 515,
 SE-751\,20 Uppsala,
 Sweden}%

\author{J.\,Bergman}
 \affiliation{%
 Swedish Institute of Space Physics,
 \AA ngstr\"om Laboratory,
 P.\,O.\,Box 537,
 SE-751\,21 Uppsala,
 Sweden}%

\author{T.\,D.\,Carozzi}
 \affiliation{%
 Astronomy and Astrophysics Group,
 Department of Physics and Astronomy,
 University of Glasgow,
 Glasgow, G12\,8QQ,
 Scotland, UK}%

\author{Ya.\,N.\,Istomin}
 \affiliation{%
 I.\,E. Tamm Theory Department,
 P.\,N. Lebedev Physical Institute,
 53 Leninsky Prospect,
 Moscow, 119991,
 Russia}%

\author{N.\,H.\,Ibragimov}
\author{R.\,Khamitova}
 \affiliation{%
  Department of Mathematics and Science,
  Research Centre ALGA: Advances in Lie Group Analysis,
  Blekinge Institute of Technology,
  SE-371\,79 Karlskrona,
  Sweden}

\begin{abstract}

We show numerically that vector antenna arrays can generate radio beams
that exhibit spin and orbital angular momentum characteristics similar
to those of helical Laguerre-Gauss laser beams in paraxial optics.  For
low frequencies ($\lesssim 1$~GHz), digital techniques can be used to
coherently measure the instantaneous, local field vectors and to
manipulate them in software.  This enables new types of experiments that
go beyond what is possible in optics. It allows information-rich radio
astronomy and paves the way for novel wireless communication concepts.

\end{abstract}


\maketitle

Classical electrodynamics exhibits a rich set of symmetries
\cite{Ribaric&Sustersic:Book:1990}, and to each Lie symmetry there
corresponds a conserved quantity
\cite{Noether:NGWG:1918,Ibragimov:JMAA:2007}.  Commonly utilized
conserved electromagnetic (em) quantities are the energy and linear
momentum, where the underlying symmetries, under Poincar\'e
transformations, are homogeneity in time and in space, respectively.

Another conserved quantity, manifesting the isotropy of space, is the em
angular momentum, whose mechanical properties were predicted
theoretically in 1909 \cite{Poynting:PRSL:1909} and demonstrated
experimentally in 1936 \cite{Beth:PR:1936}.  A collection of
nonrelativistic, spinless, classical particles with linear momenta
$\pmech_i$ has angular momentum
${\Jmech=\sum_i(\x_i-\x_0)\cross\pmech_i}$, where $\x_0$ is the moment
point.  When this system interacts with em radiation with spin angular
momentum (SAM) $\Sem$ and em orbital angular momentum (OAM) $\Lem$, the
total angular momentum, $\Jtot = \Jmech+\Jem$, where $\Jem =\Sem+\Lem =
\epsz\int\d^3\!x\;(\x-\x_0)\cross(\E\cross\B)$, is conserved
\cite{Cohen-Tannoudji&al:Book:1989}.  Hence, for a fixed $\Sem$, a
change in $\Jmech$ will result in an opposite change in $\Lem$, observed
as a rotational (azimuthal) Doppler shift \cite{Courtial&al:PRL:1998}.
This shift is distinct from the translational Doppler shift, which is a
manifestation of the conservation of linear momentum.  As shown in
Ref.~\onlinecite{Barnett:JOB:2002}, using an angular momentum flux
representation \cite{Schwinger&al:Book:1998}, it is always possible to
separate a beam $\Jem$ into $\Sem$, which depends only on the local
polarization structure, and $\Lem$, which depends on the gradient of the
fields; see also Ref.~\onlinecite{Berry:SingOpt:1998}.

During the past few decades the use of em OAM (beam vorticity) has come
to the fore in optics \cite{Allen:JOB:2002} and in atomic and molecular
physics \cite{Cohen-Tannoudji:RMP:1998}.  However, while SAM (wave
polarization), generated by proper phasing of the two legs in a crossed
dipole or by using helix antennas, has been used routinely for at least
half a century, OAM has not yet been utilized to any significant degree
in radio physics \cite{Krishnamurthy&al:Asilomar:2004} or its
applications such as radio astronomy \cite{Harwit:APJ:2003}.  The use of
radio OAM is currently being contemplated for detection of ultrahigh
energy neutrinos interacting with the Moon \cite{Stal&al:PRL:2007},
studies of radio wave interactions with the atmosphere and ionosphere
\cite{Istomin:PLA:2002,Paterson:PRL:2005,Thide&al:PRL:2005}, and radar
probing of the Sun \cite{Thide:MMWP:2004,Khotyaintsev&al:SP:2006}.

We propose to use antenna arrays for generating and detecting both SAM
and OAM in radio beams and show numerically how this works.  In such
arrays one needs access to the complete 3D vectors of the radio em field
over an area which is large enough to intersect a substantial fraction
of the radio beam.  This requires vector antennas, \eg, tripoles
\cite{Compton:IEEETAP:1981,Carozzi&al:PRE:2000}; crossed dipole antennas
will be useful for beam directions nearly perpendicular to the antenna
planes.  Using digital samplers connected directly to each vector
antenna, the local, instantaneous 3D radio field vectors themselves can
be measured coherently up to the GHz range, enabling their manipulation,
including $\Jem$ processing, entirely in software.  This is in contrast
to infrared and optical frequencies for which current detectors are
incapable of measuring first-order field quantities. There
phase-coherent down-conversion to the low-frequency radio domain
might provide a solution.

\begin{figure}[t]
\begin{minipage}{\columnwidth}
\includegraphics[width=.48\columnwidth]{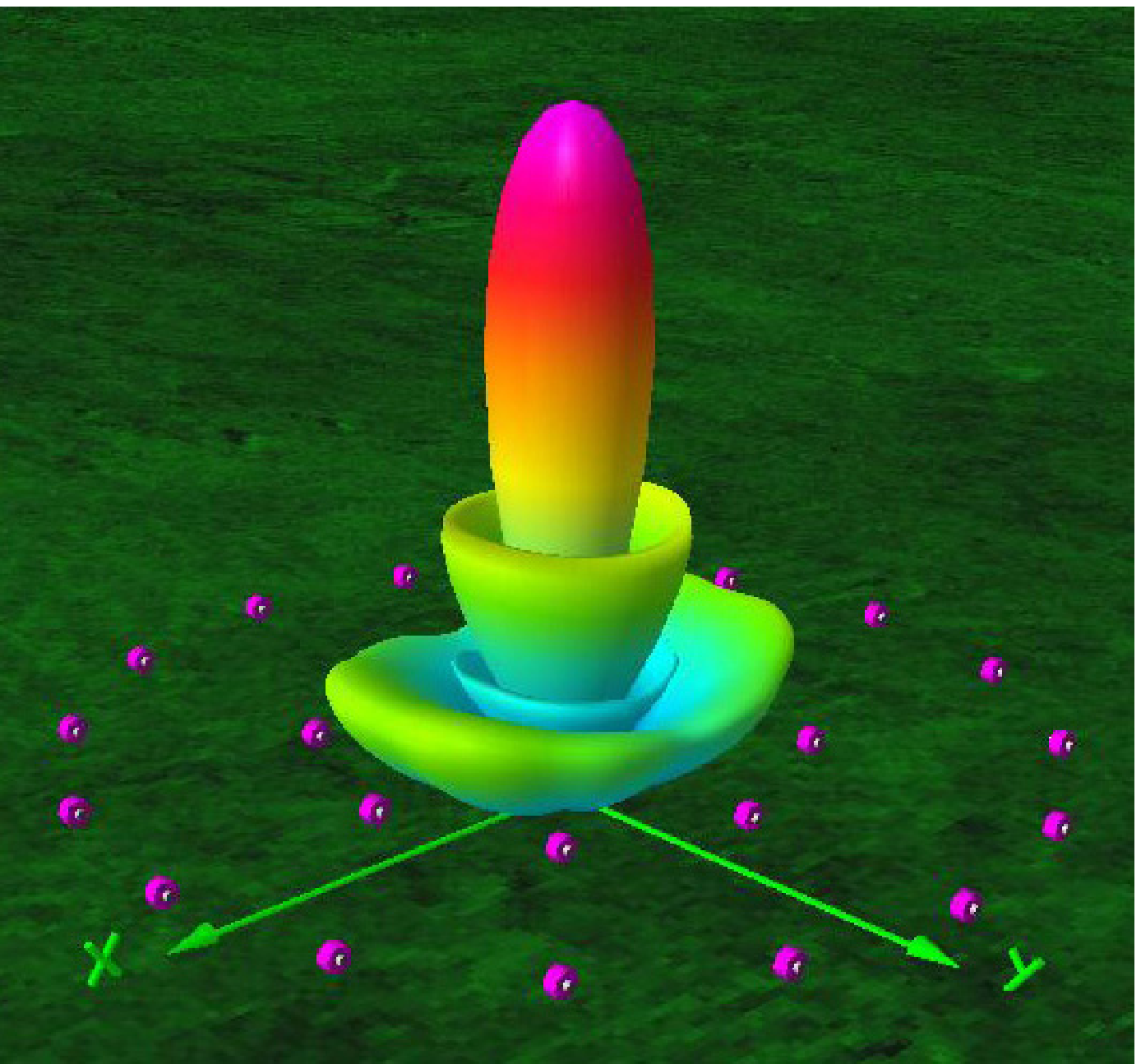}
\hfill
\includegraphics[width=.48\columnwidth]{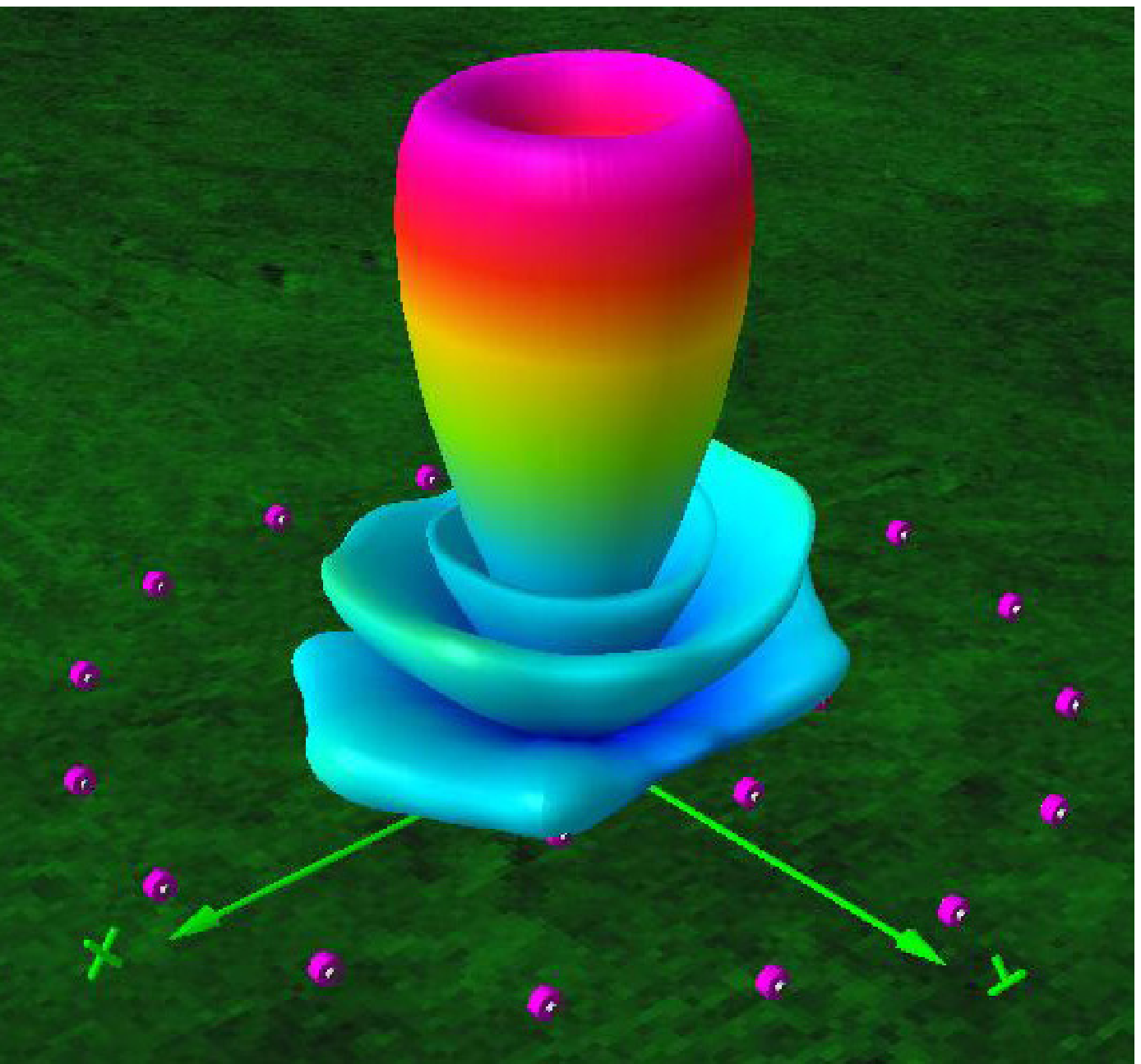}
\end{minipage}\\[1ex]
\begin{minipage}{\columnwidth}
\includegraphics[width=.48\columnwidth]{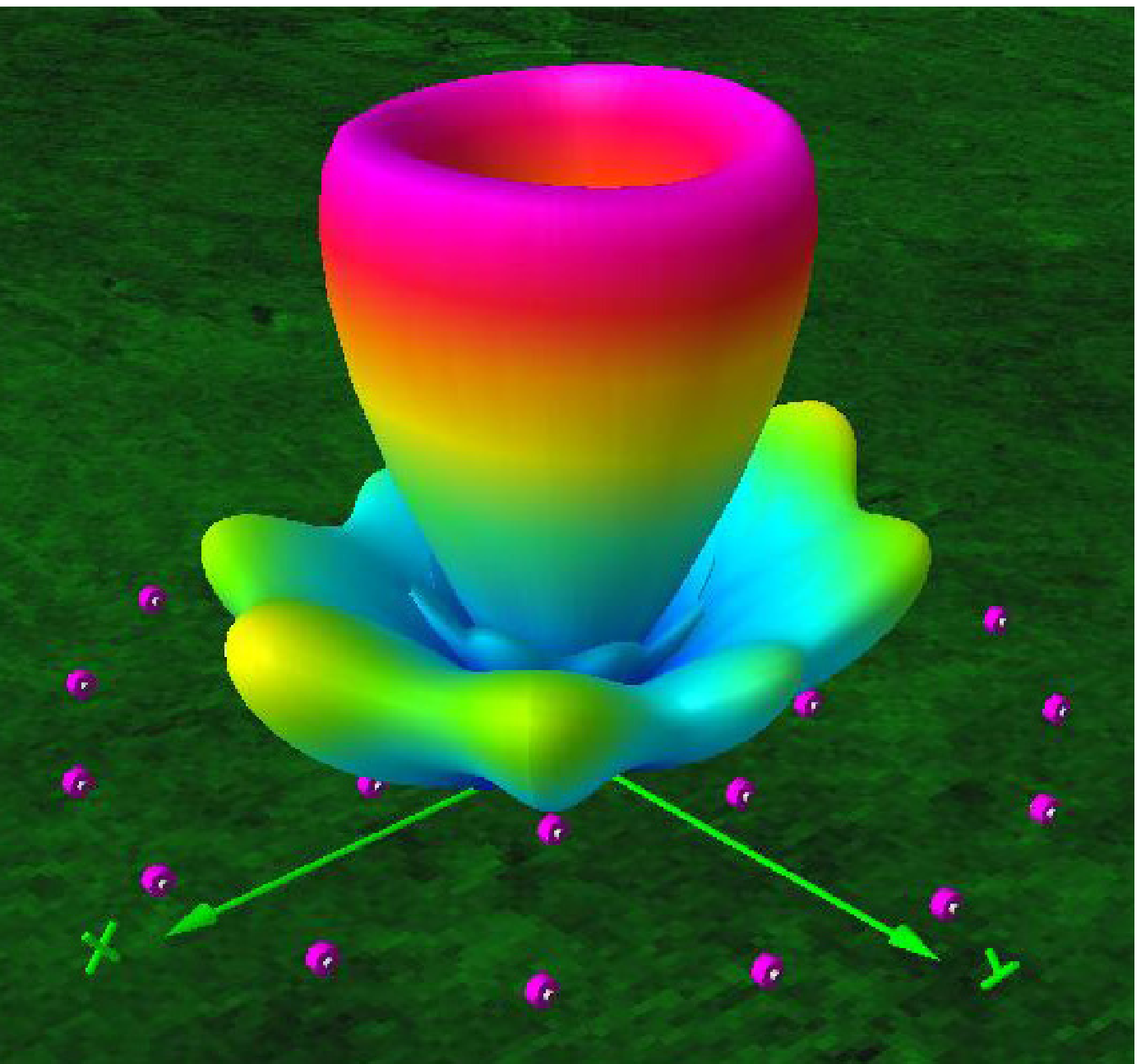}
\hfill
\includegraphics[width=.48\columnwidth]{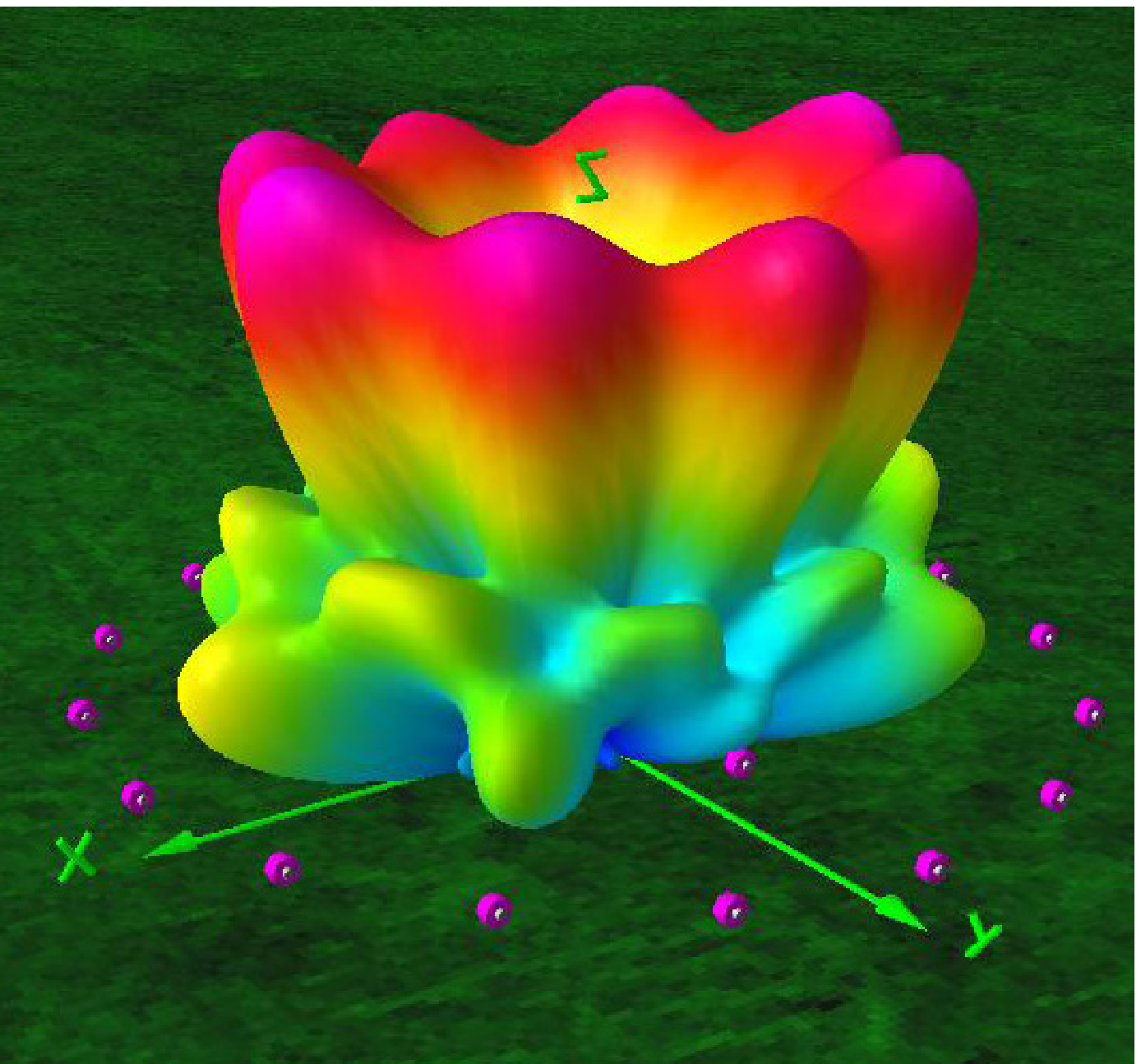}
\end{minipage}
\caption{%
(color online).
Radiation patterns for radio beams generated by one circle of 8 antennas
and radius $\lambda$ plus a concentric circle with 16 antennas and
radius $2\lambda$; all antennas are $0.25\lambda$ over the ground.  Notice
the influence of $l$ on the radiation pattern.  Here $l=0$ (upper left),
$l=1$ (upper right), $l=2$ (lower left), and $l=4$ (lower right).}
\label{fig:patterns}
\end{figure}

\begin{figure}[t]
\begin{minipage}{\columnwidth}
\includegraphics[width=.45\columnwidth]{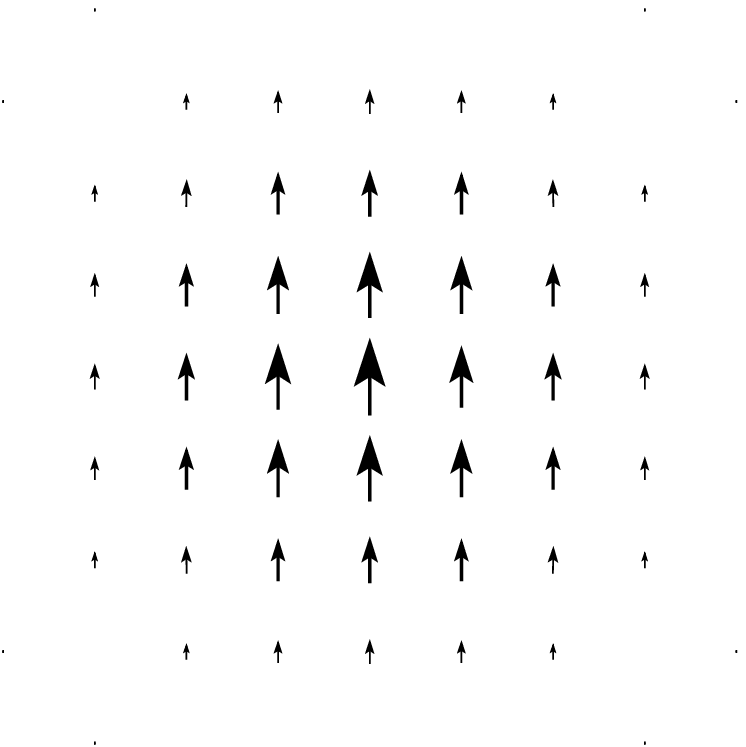}
\hfil
\includegraphics[width=.45\columnwidth]{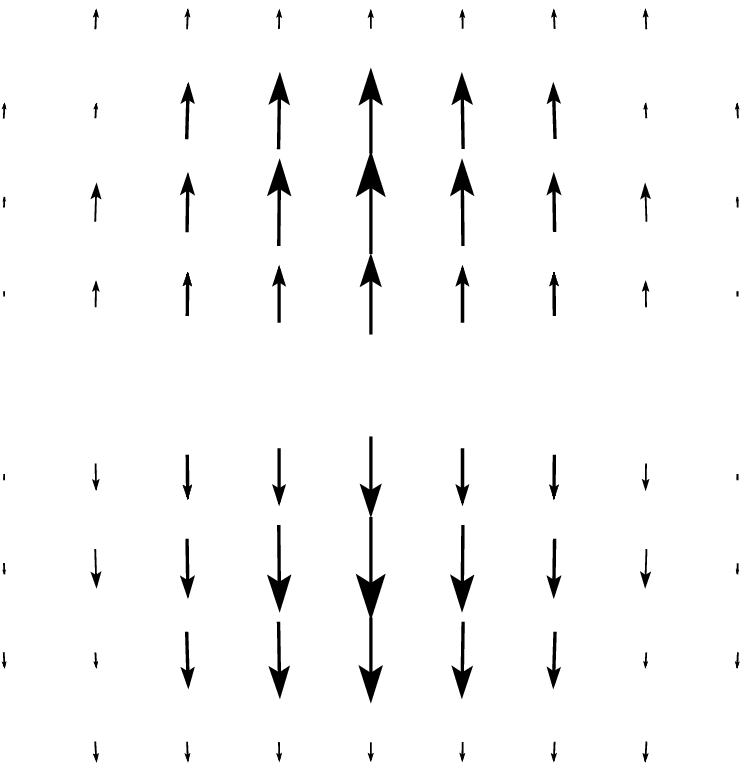}
\end{minipage}\\
\begin{minipage}{\columnwidth}
\includegraphics[width=.45\columnwidth]{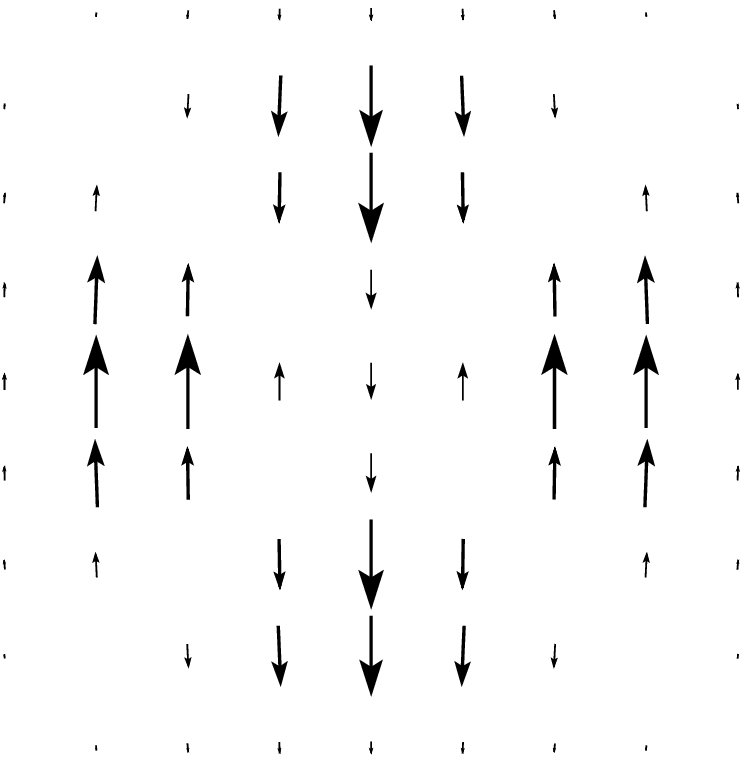}
\hfil
\includegraphics[width=.45\columnwidth]{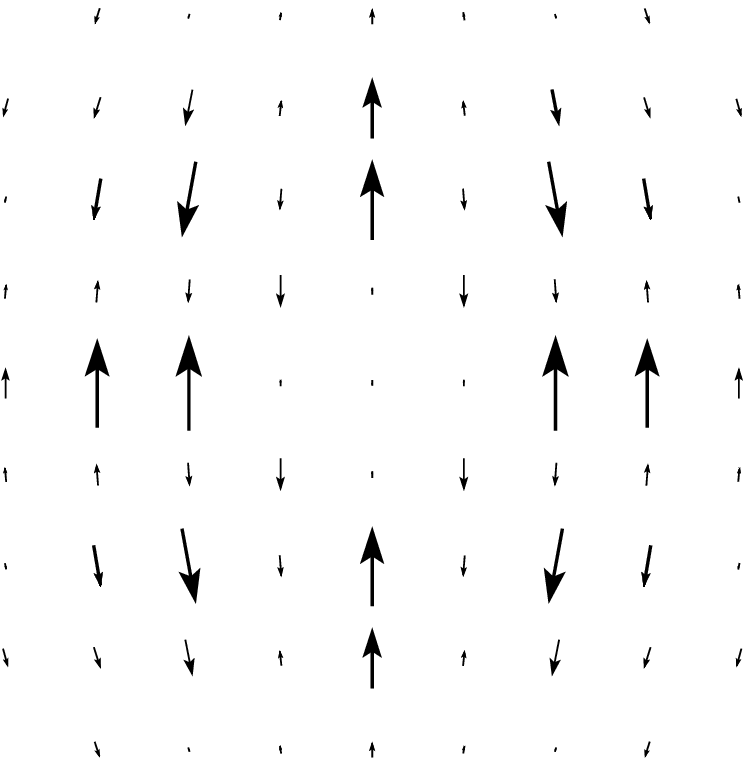}
\end{minipage}
\caption{%
Samples of instantaneous electric field vectors $\E$ across the main lobes
of the beams in Fig.~\ref{fig:patterns} (same $l$ values and plotting
order). The size of an arrow is linearly proportional to the local $|\E|$.
As expected for OAM carrying beams, the phase of the em field changes
by $l2\pi$ for a full turn around the beam axis.
}
\label{fig:Efields}
\end{figure}

Lasers often use Laguerre-Gaussian (LG) modes \cite{Siegman:Book:1986}
in which the phases of the electric and magnetic vector fields in a
plane perpendicular to the beam axis have an $l\varphi$ dependence where
$l$ is an integer and $\varphi$ is the azimuthal angle.  This means that
for $l\neq0$ the phase fronts of LG modes are not planar but helical.
As shown in Ref.~\onlinecite{Allen&al:PRA:1992}, this implies that LG
beams carry an OAM of $l\hbar$ per photon.  In the paraxial
approximation, the LG modes form a complete basis set for light beams
\cite{Siegman:Book:1986}.

\begin{figure}
\begin{minipage}{\columnwidth}
\includegraphics[width=0.32\columnwidth]{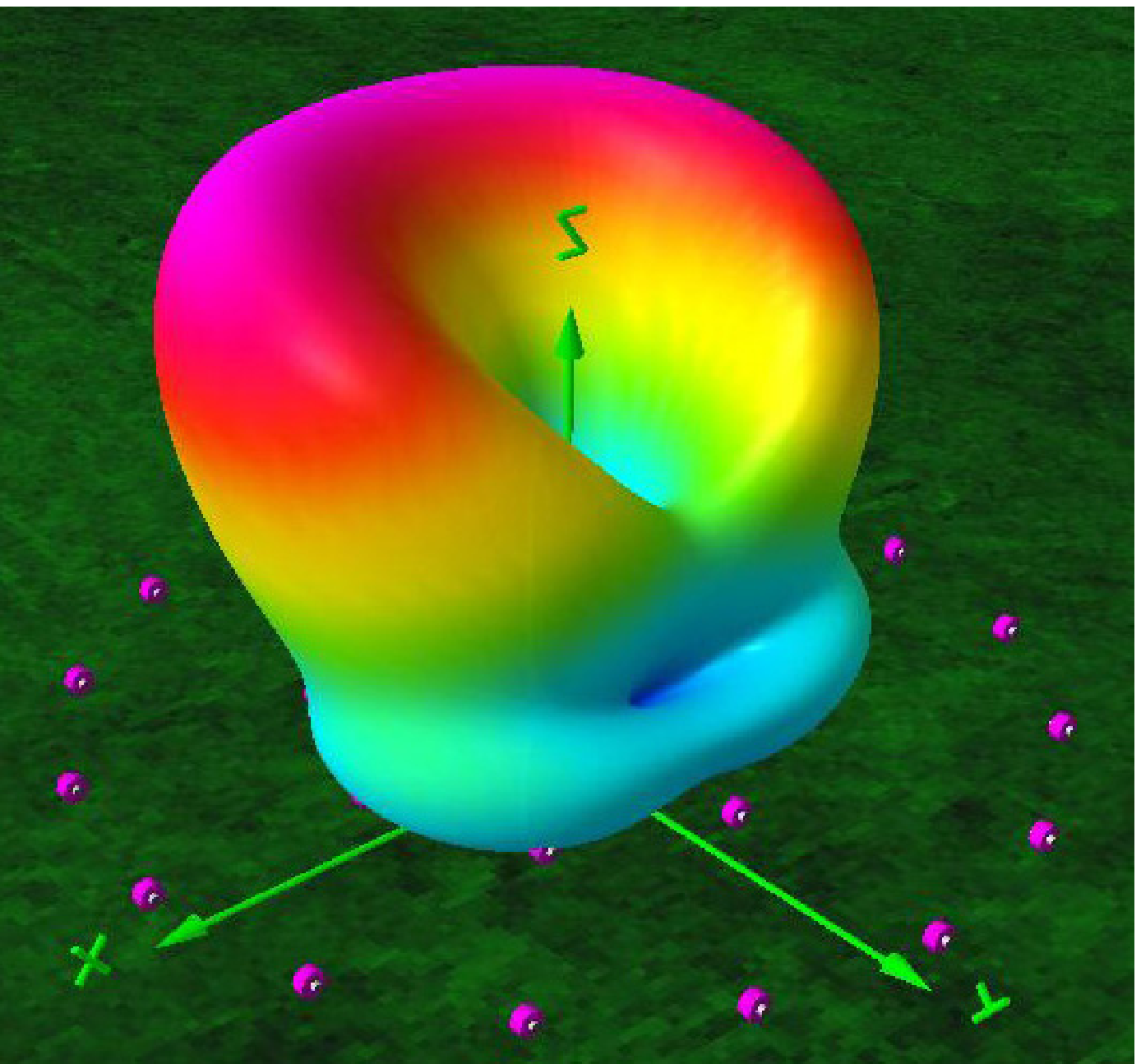}
\includegraphics[width=0.32\columnwidth]{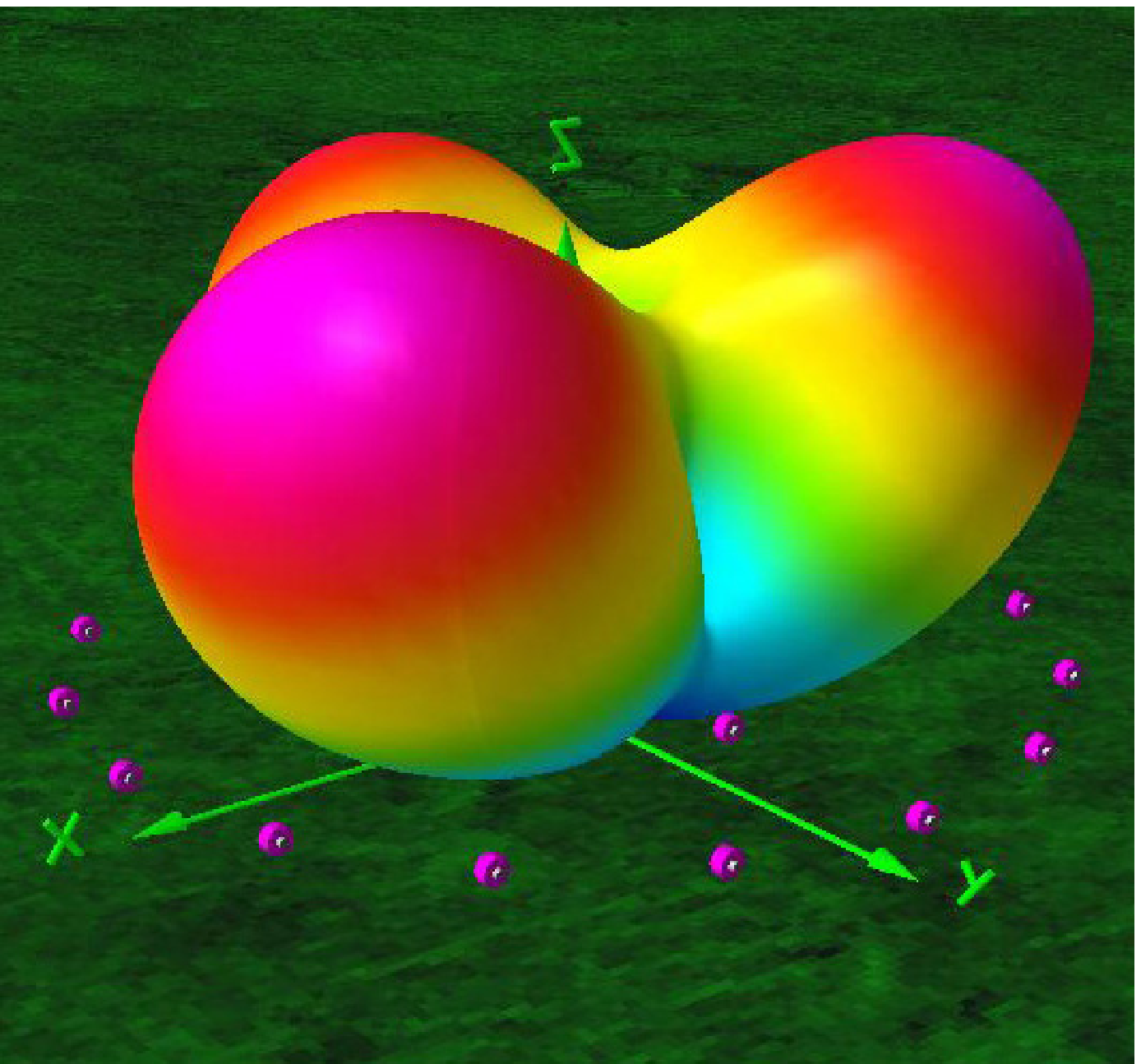}
\includegraphics[width=0.32\columnwidth]{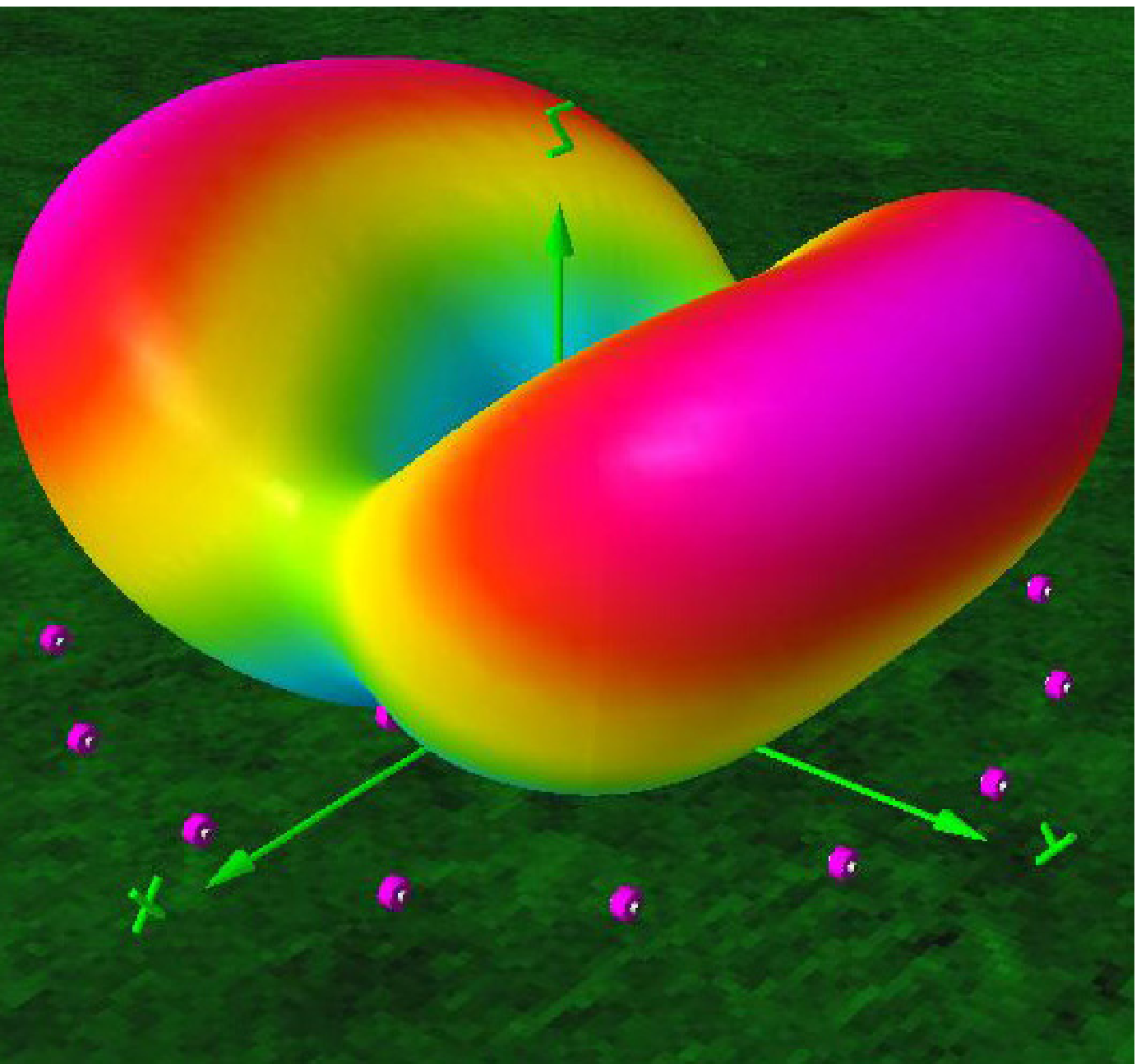}
\end{minipage}
\begin{minipage}{\columnwidth}
\includegraphics[width=0.32\textwidth]{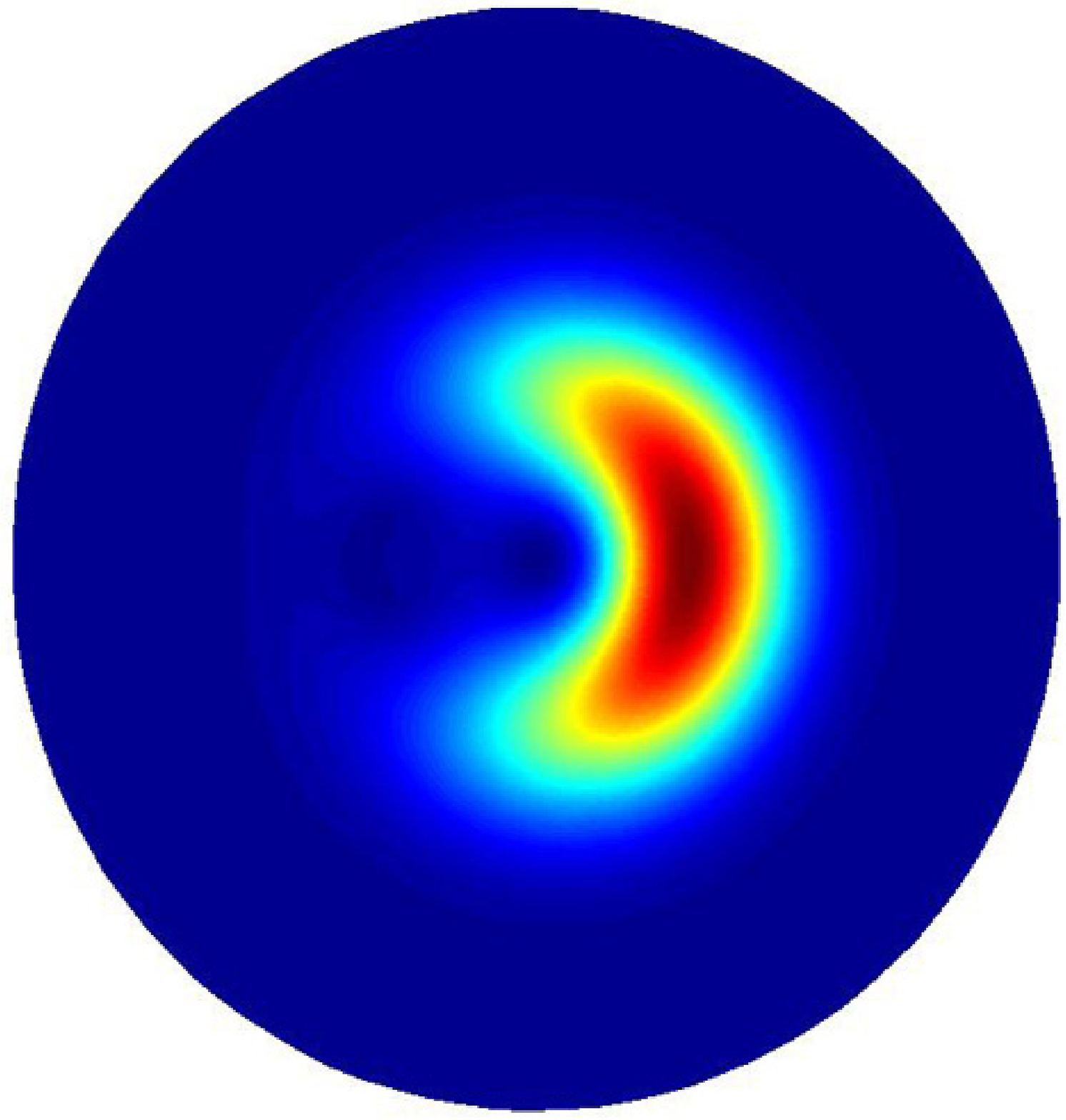}
\includegraphics[width=0.32\textwidth]{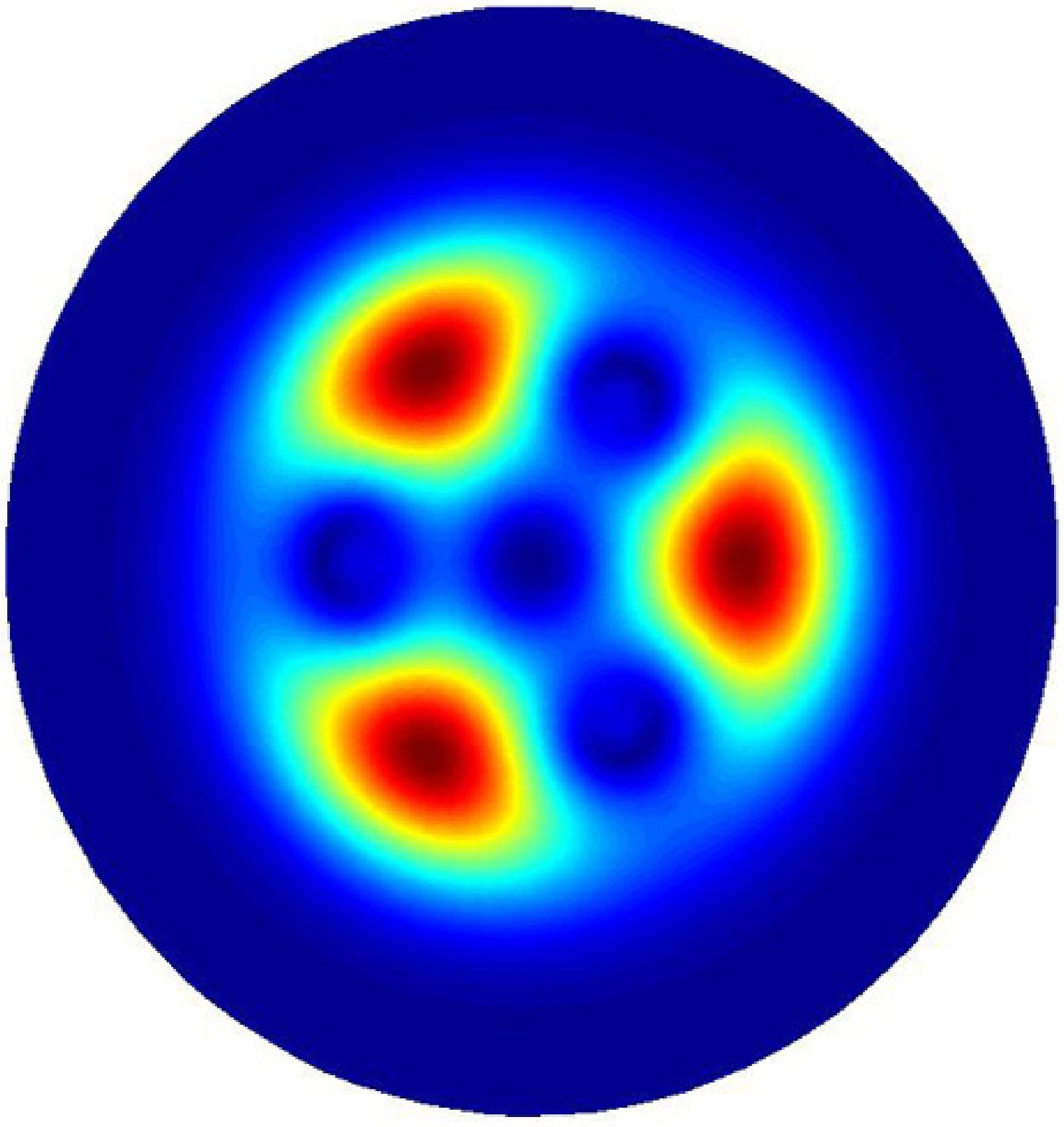}
\includegraphics[width=0.32\textwidth]{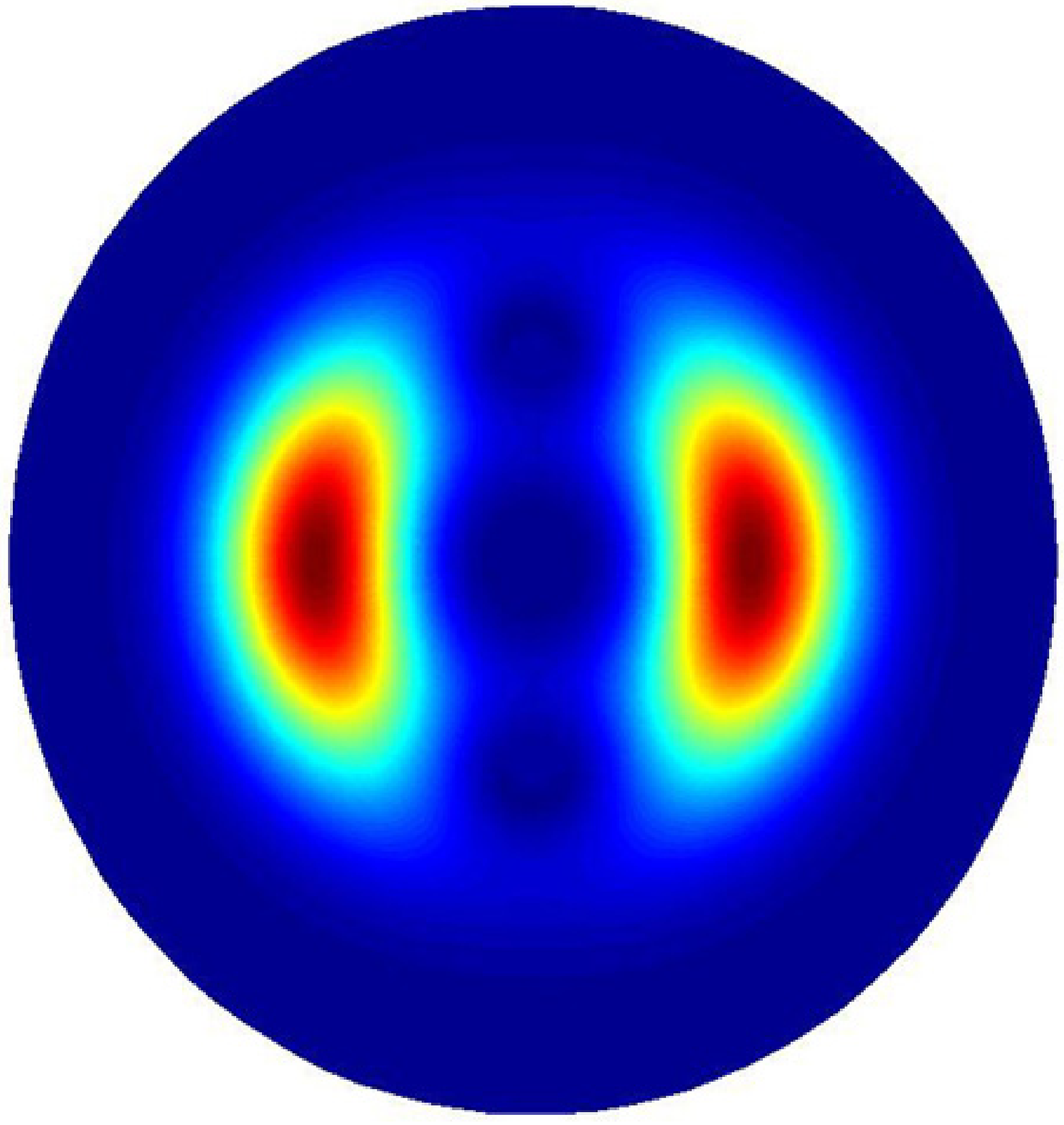}
\end{minipage}
\caption{%
(color online).
Beams obtained by superimposing two different OAM states.  The upper
three panels show the radiation patterns for the antenna array, and the
lower three panels show the corresponding intensity patterns, head on,
calculated for Laguerre-Gaussian beams.  The leftmost are for $l_1=1$
and $l_2=2$, the middle ones are for $l_1=1$ and $l_2=4$, and the
rightmost are for $l_1=2$ and $l_2=4$.  Notice the good agreement
between the patterns obtained with the antenna array model and the
paraxial LG beam model.
}
\label{fig:MIX}
\end{figure}

\begin{figure}
\begin{minipage}{\columnwidth}
\includegraphics[width=0.49\textwidth]{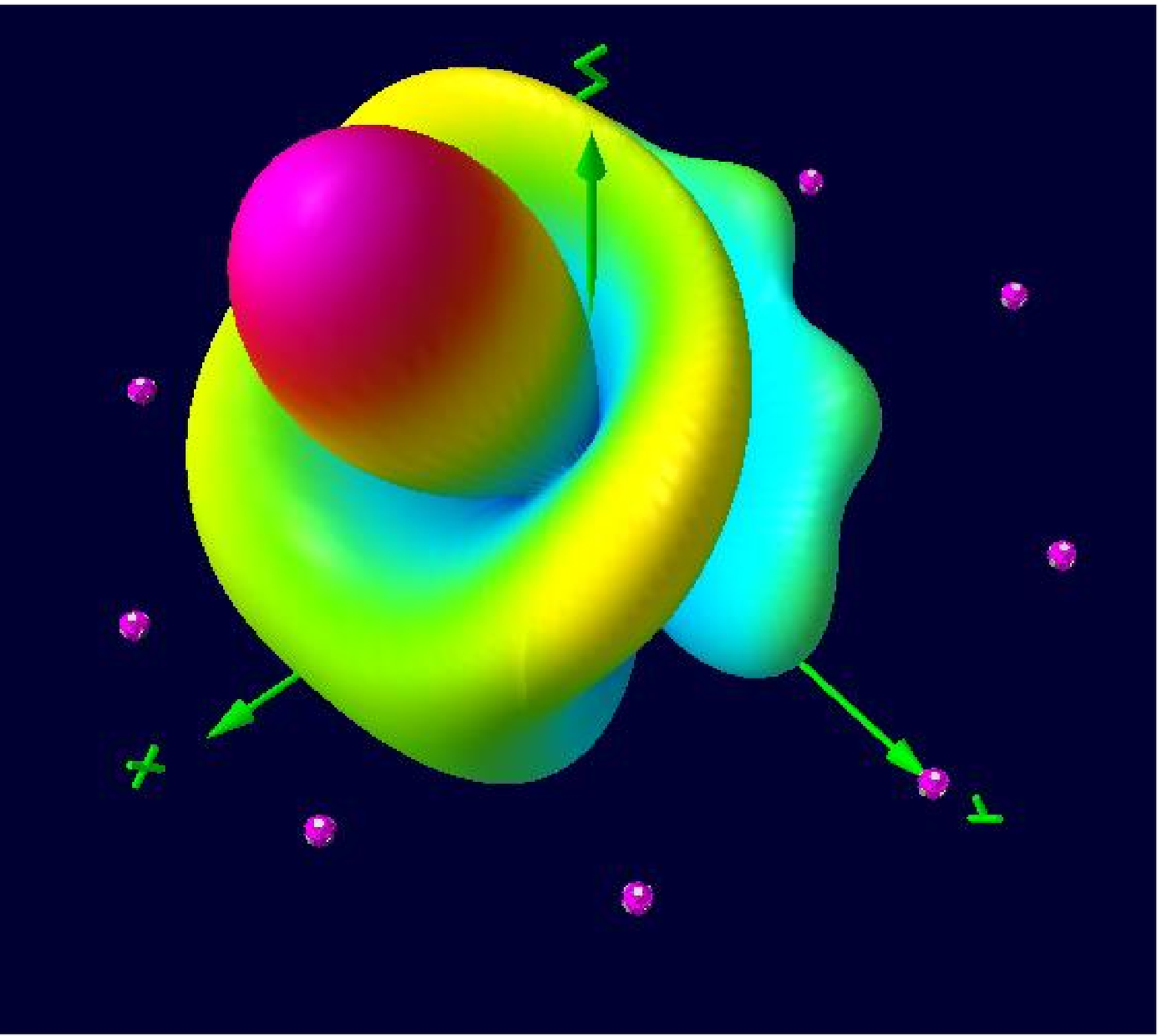}
\hfill
\includegraphics[width=0.49\textwidth]{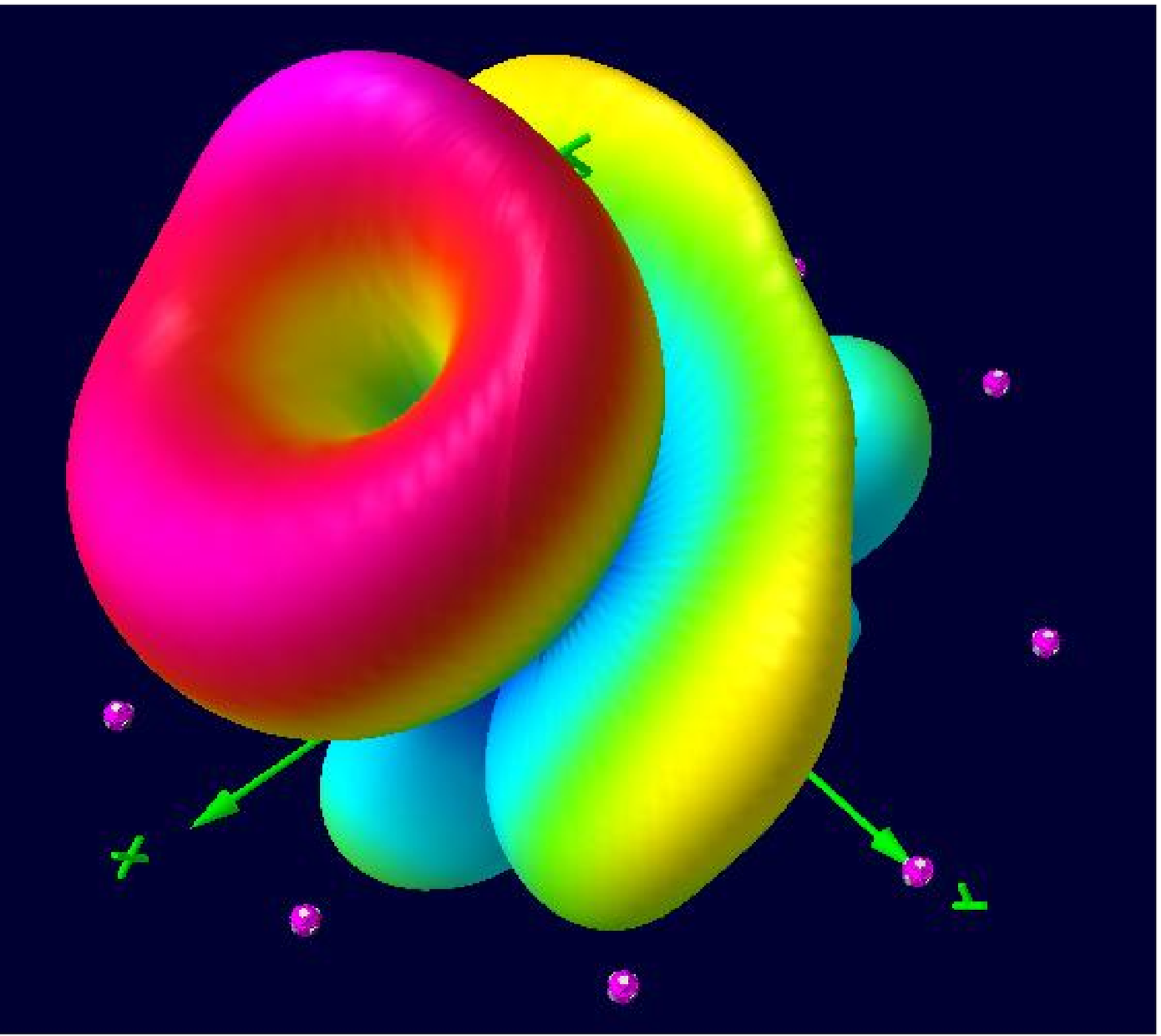}
\end{minipage}
\caption{%
(color online).
Radiation patterns for a circularly polarized beam propagating
obliquely ($\theta=25\deg$) with $l=0$ (left) and $l=1$ (right),
generated by phasing the individual elements of a ten-tripole array in
free space. This illustrates that with a tripole array it is possible
to control electronically both the beam direction and $l$. This is
not possible with arrays of single or crossed dipoles. Note that for
$l\neq0$, there will be an on-beam-axis minimum which can be useful to
block out a bright object when observing faint surrounding objects
\cite{Swartzlander:OL:2001}, \eg, in the solar corona
\cite{Khotyaintsev&al:SP:2006}.}
\label{fig:tripoles}
\end{figure}

A pure OAM state radio beam of frequency $\w$ and energy $H$ has a beam
axis OAM component ${\Lzem=lH/\w}$ and the fields have an azimuthal
phase dependence of $\exp(\i l\varphi)$, where $l$ is an integer as in
an LG beam.  In order to study the possibility of using OAM in radio, we
consider an antenna array and assume for simplicity that each antenna is
located equidistantly along the perimeters of circles (see
Fig.~\ref{fig:patterns}).  The antennas are fed the same signal, but
successively delayed relative to each other such that after a full turn
around the antenna array axis, the phase has been incremented by
$l2\pi$.  The far field intensity patterns, calculated with the software
package NEC2 \cite{NEC2} which solves Maxwell's equations for a given
set of antenna currents, are displayed in Figs.~\ref{fig:patterns},
\ref{fig:MIX}, and \ref{fig:tripoles} and are very similar to those
obtained in paraxial optics.  Fig.~\ref{fig:Efields} displays the
instantaneous $\E$ field vectors across the main lobes for different
$l$.  Theory predicts that $j=l+s=\w\Jzem/H$ \cite{Allen&al:PRA:1992}.
Table~\ref{tab:wJzH} lists the results obtained for $l=0,1,2,3$ and
$s=-1$ for a radio beam generated by an antenna array.  The agreement is
excellent.

The superposition of two coaxial OAM states was modeled numerically
by using two concentric antenna rings with different radii.  In the
inner ring we increase the overall phase between successive antennas by
$2\pi l_1/n_1$.  In the outer ring we use the phase increment $2\pi
l_2/n_2$.  The variables $n_1$ and $n_2$ are the number of antennas in
each ring, respectively, and $l_1$ and $l_2$ are the phase increment
factors corresponding to the OAM number $l$ in the LG beams.  The
resulting antenna radiation patterns are shown in Fig.~\ref{fig:MIX} for
three different combinations of $l_1$ and $l_2$.  These patterns would
be difficult to synthesize without resort to the OAM technique.  For
comparison, intensity patterns for OAM carrying LG beams for the same
$l_1$ and $l_2$ combinations are also shown.

\begin{table}[b]
\caption{Scaling of $\w\Jzem/H$ as a function of $l$ for a right-hand
circular polarized beam ($s=-1$) formed by a ring array of 10 crossed dipoles.
Array radius $D=\lambda$, antennas $0.1\lambda$ over perfect ground,
polar angle $\theta=0$.}
\label{tab:wJzH}
\begin{tabular*}{1.0\columnwidth}{c@{\extracolsep{\fill}}ccc}
\hline
\hline
$l$ & $s$ & $j=l+s$ & $\w\Jzem/H$ \\
\hline
0 & -1 & -1 & -1.019  \\
1 & -1 &  0 & -0.022  \\
2 & -1 &  1 &  0.971  \\
3 & -1 &  2 &  1.81   \\
\hline
\hline
\end{tabular*}
\end{table}

In the general (linear) case, the total OAM in a beam is a superposition
of several OAM states.  This superposition can be decomposed into pure
OAM states via a discrete Fourier transform.  In particular, one has to
integrate the complex field vector weighted with $\exp(-\i l\varphi)$
along a circle around the beam axis.  Since there will be only a finite
number of antennas along the integration path, there is an upper limit
on the largest OAM number that can be resolved.  Namely,
$|l|<$\#\{antennas on a circle around the beam axis\}$/2$.  We have
assumed that the beam axis is centered on the antenna array, but this is
not necessarily true.  The beam axis is determined by the emitter only,
not by the receiving antenna array.  Therefore an incoming radio beam
might not overlap the antenna array perfectly so that only an asymmetric
spatial part of the beam can be analyzed.  A disadvantage in this case
is that we do not have exact information of the phase along the whole
circle around the beam axis.  We have to extrapolate from accurate
measurements of the field vectors within the finite size antenna array
to the field vectors around the beam axis.  This extrapolation yields an
uncertainty $\Delta{}l>R\Delta\varphi/D,(R\gg D)$, where
$\Delta\varphi$ is the smallest phase difference in radians that can be
resolved, $D$ is the diameter of the antenna array, and $R$ is the
distance from the beam axis to the antenna array.  Since $l$ is integer
valued, the uncertainty does not matter as long as it is less than
$1/2$.  While losing information about the individual OAM states, it is
possible to measure estimated larger OAM numbers up to
$|l|<$\#\{antennas along a circle segment of length $D$\}$\pi R/D,(R\gg D)$.

The Poynting vector of a radio beam with OAM has a helical phase structure
and spirals around the main beam axis with a pitch
angle $\alpha_l=\arctan(\lambda l/2\pi R)$.  This angle can be resolved if
$l$ can be resolved, \ie, if $\Delta l<1/2$.  In this case one observes
multiple images of a single point source where each image corresponds
to a pure OAM state $l$.  If $l$ cannot be resolved, the multiple images
blend together yielding a smeared spot.  Conversely, if the individual OAM
states are resolved, the smeared spots are resolved as multiple images
(one for each $l$), enabling self-calibration techniques that sharpen
the radio image via the use of OAM.  

Inserting typical values of $D=100$\,km,
$\Delta\varphi=2\pi\times1^\circ/360^\circ$, into
$\Delta{}l>R\Delta\varphi/D$, and requiring that ${\Delta l<1/2}$, one
obtains $R\sub{max}=3000$\,km for which the OAM can be resolved.  It
should therefore be possible to probe OAM processes in the Earth's
ionosphere and lower magnetosphere and also to actively induce plasma
vorticity and strong toroidal electric currents there
\cite{Istomin:PLA:2002}.  In particular, a radio beam that interacts
with a turbulent medium will carry information on the vorticity of this
medium allowing for remote radio imaging of the turbulence
\cite{Paterson:PRL:2005}.  But, it is not likely that radio beams from
distant astronomical objects are so narrowly focused that they can be
probed for OAM unless the array extends into space.  On the other hand,
if the ionosphere adds OAM to a beam from a distant source, this added
OAM can be measured and compensated for.

Should, however, the radio emitting object rotate fast and have sharp
discrete lines in its emission spectrum, the angular momentum of the
emitted em waves can be measured indirectly via shifts and splittings of
the spectral lines.  The shifts result from the rotational Doppler
effect $\w-\w_0=(l+s)\Omega_\parallel$ where $\Omega_\parallel$ is the
projection of the rotation frequency onto the wave vector $\k$
\cite{Courtial&al:PRL:1998}.  Decomposing into pure spin states, the
discrete emission spectrum will be decomposed into one for $s=+1$ and
one for $s=-1$.  These two spectra should almost coincide in their
spectral lines, except for an overall shift.  The relative overall shift
between the two spectra is equal to twice the rotational frequency of
the emitter.  Once $\Omega_\parallel$ has been read off, one can search
for spectral lines that are separated exactly by $\Omega_\parallel$ (and
integer multiples thereof).  Each of these spectral lines corresponds to
a specific OAM state.

When pointing the beam in a direction other than the $z$ axis
(orthogonal to the plane of the dipoles), vector sensing antennas such
as tripoles are much preferred.  They have the advantage that the three
orthogonal antenna currents can be viewed as the $x$, $y$, and $z$
components of a composite antenna current vector $\j(t,\x)$ which, under
software control, can be rotated into any given direction in space, in
the receive case even after the fact.  To generate off-zenith em beams
with non-zero $l$ values, the tripole is indispensable since one cannot
control the direction of the beam with linear dipoles or crossed dipoles
and still maintain the desired $l\varphi$ phase dependence; see
Fig.~\ref{fig:tripoles}.  This is because the phase difference between
the elements is used to point the beam.  The only way to overcome this
limitation would be to mechanically rotate the antenna array.  Such an
arrangement would be infeasible.

In summary, we have shown by theoretical and numerical modeling that a
number of the photon SAM and OAM signatures observed in paraxial optics
can be generated in the radio frequency range where, for low enough
frequencies ($\lesssim1$~GHz), modern digital radio techniques can be
used to measure coherently the 3D electric (and magnetic) vectors in the
beams.  This opens up new kinds of fundamental electromagnetic
radiation experiments.  Furthermore, minor changes of the design of the
large low-frequency multi-array radio telescopes coming on line (LOFAR
\cite{LOFAR}) or in the planning stage (SKA \cite{SKA}), would enable
them to utilize not only SAM but also OAM, which would increase their
resolution, sensitivity, interference tolerance, and overall usefulness.
The LOIS (LOFAR Outrigger in Scandinavia) Test Station in southern
Sweden has already implemented the radio OAM techniques and is currently
being set up for the first proof-of-concept experiments
\cite{Thide:MMWP:2004}.

Since information can be encoded as OAM states \cite{Gibson&al:OE:2004}
that span a much larger state space than the two-state SAM space
\cite{Leach&al:PRL:2002}, radio OAM techniques hold promise for the
development of novel information-rich radar and wireless communication
concepts and methodologies. Furthermore, it is conceivable that
signatures related to radio OAM or other conserved quantities in Dirac's
symmetrized form of the Maxwell-Lorentz em theory might provide clues
on the existence of magnetic monopoles \cite{Ibragimov&al:JMP:2007}.
Finally, it should be mentioned that invariants are very useful for
assessing the stability and robustness of numerical simulation codes.

\begin{acknowledgments}
The authors thank Sir Michael Berry, Bruce Elmegreen, Bengt Eliasson,
Erik B.~Karlsson, Dan Stinebring, and Willem Baan for elucidating
discussions and helpful comments.  We gratefully acknowledge the
financial support from the Swedish Governmental Agency for Innovation
Systems (VINNOVA).
\end{acknowledgments}

\bibliographystyle{apsrev}
\bibliography{oam}

\end{document}